# Empirical correlation of the surface tension versus the viscosity for saturated normal liquids

**Xia Li [1], Jianxiang Tian [1,2,*], A. Mulero [3]**

[1]Shandong Provincial Key Laboratory of Laser Polarization and Information Technology

Department of Physics, Qufu Normal University, Qufu, 273165, P. R. China

[2]Department of Physics, Dalian University of Technology, Dalian, 116024, P. R. China

[3]Department of Applied Physics, University of Extremadura, Badajoz 06006, Spain

[*]Corresponding author, E-mail address: jxtian@dlut.edu.cn

**Abstract**

In 1966 Pelofsky proposed an empirical linear correlation between the natural logarithm of the surface tension and the reciprocal viscosity, which seems to work adequately for a wide range of fluids. In particular, it has been shown that it is useful in the case of n-alkanes and their binary and ternary mixtures. More recently however, it has been found not to work for several ionic liquids unless the reciprocal viscosity is raised to a power. The exponent of this power was fixed to be 0.3, at least for the studied ionic fluids. In the present work, the performance and accuracy of both the original Pelofsky correlation and the modified expression including the exponent are studied for 56 non-ionic fluids of different kinds over a broad range of temperatures. Also, the temperature range is delimited for which each expression reproduces the surface tension values with average absolute deviations below 1%. The needed coefficients are given for both the broad and the delimited temperature range for each expression. Unfortunately, the results show that the value of the exponent in the modified Pelofsky expression is substance-dependent for the normal fluids studied.

# 1. Introduction

Surface tension and viscosity are two properties of fluids which are different in nature but whose values need to be known for a wide variety of industrial and physicochemical processes. Surface tension affects important stages in such production processes as catalysis, adsorption, distillation, and extraction, and viscosity is important in processes involving a flow of fluids, such as the use of lubricants. The two properties have been extensively studied for normal fluids, and this interest continues (see [1-5] for instance). In particular, the broad range of applications of ionic fluids has led to burgeoning interest in research on their surface tension and viscosity [6-10].

While surface tension and viscosity can be measured with high accuracy for low and moderate temperatures, for high temperatures predictions using computer simulations such as Monte Carlo methods are commonly required [3, 11]. For some fluids, one of these properties may be more easily measured than the other for certain temperature ranges. It is therefore interesting to try to establish some relationship between the two properties. Such a relationship could also be used to test the validity of the measured data, since deviations may be due to experimental error [12]. Indeed, since both properties are related to the intermolecular potential energy, one might expect there to be some theoretical correlation between the two, although no such link has yet been established.

In 1966, Pelofsky [13] proposed an empirical relationship between the natural logarithm of surface tension and the inverse of viscosity (usually termed the fluidity). Pelofsky's empirical expression (which we shall denote here as the P correlation) can be applied to both the organic and inorganic phases of pure and mixed components [13]. Two adjustable coefficients are needed, whose values may depend in the temperature range being considered. The P correlation was later modified by Schonhorn [14] who introduced a correction into the second term of the right-hand side of the expression to fulfil the requirement that, at the critical point, the surface tension goes to zero while the viscosity tends to a small constant value. This modification introduces new coefficients, and has not subsequently been used.

More recently, Queimada *et al.* [12] checked the use of the P correlation for pure compounds and mixtures of n-alkanes, and found adequate results in all cases. The temperature ranges they considered were, however, fairly narrow. Thus, for the pure compounds, the differences between the maximum and minimum temperatures were from about 60 K to about 110 K, and the maximum temperatures were not close to the critical point. Indeed, the authors themselves observed that the use of the P correlation for temperatures near the critical point could yield very inaccurate results.

Ghatee *et al.* [10] applied the P correlation to ionic fluids. In particular, they studied quaternary ammonium, imidazolium-, pyridinium-, pyrrolidinium-, and nicotinium-based ionic liquids. They found that it was necessary to modify the P correlation slightly by introducing an exponent into the viscosity term (we shall denote this hereafter as the modified P, or MP, correlation). They initially treated this exponent as an adjustable coefficient, but then found that its value could be fixed at 0.3 without any significant loss of accuracy. As was also the case for Pelofsky's study, the data then available were not as accurate and extensive as presently, so that one does not know the real applicability and accuracy of their proposed correlation. Moreover, since the MP expression was applied only to ionic fluids, it would be interesting to examine whether it works adequately for normal fluids and whether there is an almost universal (*i.e.*, substance independent) value for the exponent.

In the present work, we therefore studied the performance and accuracy of both the original P and the modified MP relationships for 56 normal fluids of different kinds. In order to have as extensive a range of data as possible from the same source, we considered only those fluids for which the NIST Web Book [15] offers the values of the surface tension and viscosity over a wide range of temperatures. We therefore included simple fluids (such as argon and carbon dioxide), simple alkanes, refrigerants, and some other substances such as water and ammonia.

The absolute average deviations (AADs) of the predicted the surface tension values vary greatly from one fluid to another, but in general are clearly lower with the MP correlation. While this was to be expected since MP includes an extra adjustable parameter, the degree of improvement in the prediction is very significant for 61% of the fluids considered. Nonetheless, for most of the fluids the AAD values are greater than 1%. Therefore, for every correlation we delimited the temperature ranges for which the AAD < 1%. The coefficients required are given for both the entire and the delimited temperature ranges for every expression.

In Sec. 2 we present both the P and the MP correlations, and in Sec. 3 the results and a discussion. The final conclusions are summarized in Sec. 4.

## 2. Correlation models

The Pelofsky empirical correlation is [13]

$$\ln \sigma = \ln A + \frac{B}{\eta} \qquad (1)$$

where $\sigma$ and $\eta$ are the surface tension and viscosity, respectively, and $A$ and $B$ are substance-dependent coefficients that have to be calculated for every fluid.

The modified Pelofsky expression proposed by Ghatee *et al.* [10] was initially applied to a set of ionic liquids. From considering the non-linear behaviour of the fluidity ($\eta^{-1}$) with temperature, they concluded that it could be appropriately correlated by using the modified expression [10]:

$$\left(\frac{1}{\eta}\right)^{\phi} = a + bT \quad , \tag{2}$$

where $a$ and $b$ are substance dependent coefficients and $\phi$ is a characteristic exponent. They examined 49 ionic liquids, and found that Eq. (2) fits the temperature-dependent viscosity of those liquids quite accurately with just a single universal exponent $\phi = 0.300$.

Since the surface tension of fluids, including ionic liquids, is almost perfectly linear with temperature, and using the P expression of Eq. (1), Ghatee *et al.* proposed the MP relationship between the two properties as follows [10]:

$$\ln \sigma = \ln C + D\left(\frac{1}{\eta}\right)^{\phi} \tag{3}$$

where $C$ and $D$ are substance dependent constants, and $\phi$ takes the universal (for the ionic liquids studied in [10]) value as 0.3. They report an overall AAD value of only 0.098%, which clearly demonstrates the accuracy of Eq. (3) for the cases they considered.

As noted in the Introduction, it is interesting to determine whether the P correlation works appropriately for common fluids, and what its accuracy is when it is applied to predict the surface tension values. Since the P expression includes the natural logarithm of this property, and since surface tension values differ greatly from one substance to another, a good linear relationship does not necessarily mean that low AAD values will be obtained the corresponding predictive expression. One needs to consider both the applicability of the P expression (*i.e.*, whether such a linear relationship can be established) and its accuracy in predicting surface tension values. We therefore calculate here $R^2$ (*i.e.*, the square of the linear correlation coefficient which denotes the validity of the linear correlation between $\ln \sigma$ and $1/\eta$) and AAD for every fluid in as wide a temperature range as possible. We perform the same study for the case of the MP expression, but in this case also examining whether or not any improvement observed is significant with respect to the original expression which has one less adjustable coefficient. We also delimit the temperature ranges in which each expression gives excellent accuracy (AAD < 1%). These calculations were done for 56 fluids, including inert liquids such as argon, polar liquids such as water, non-polar fluids such as carbon dioxide, and several hydrocarbons and refrigerants.

## 3. Results

As indicated above, we used data obtained from the NIST Web Book [15] because they are sufficiently accurate and are publicly and straightforwardly available. The data on the saturation curves are available for a certain temperature range, usually between the triple and the critical points, and limited to a maximum of 201 data points. Since the surface tension is defined as zero at the critical point, we excluded this datum, so that the default number of data for each fluid was 200. Nevertheless, we found that for certain fluids the surface tension and viscosity data are not both available for some low or high temperature ranges. For instance, in the case of the refrigerant R11 there are data for the surface tension near both the triple and the critical points, but not for the viscosity, so that, in this case, the final number of data used was only 175.

We note that, according to Mulero *et al.* [16], in the cases of ammonia and neon the presently available NIST surface tension data are inadequate, so that we used as a proxy the new correlation recommended in that work to generate the appropriate values.

The 56 fluids studied and their critical point temperatures are listed in Table 1 in alphabetic order for three kinds of substances: refrigerants, hydrocarbons, and other common fluids. The data used start at the temperature $T_0$, which is just the triple point temperature except for the fluids marked with an asterisk, and finish at the temperature $T_f$, which is automatically selected by the software in the NIST Web Book and is as near to the critical point temperature as possible. In the table, $N$ is the number of data used for every fluid, and $\phi$ is the value of the exponent in the MP expression (which is exactly 1 in the case of the P correlation).

For every fluid, we checked the performance of the two correlations by calculating the linear coefficient of determination $R^2$ as the square of the following expression:

$$R = \frac{\sum_{i=1}^{n}(x_i - \bar{x})(y_i - \bar{y})}{\sqrt{\sum_{i=1}^{n}(x_i - \bar{x})^2 \sum_{i=1}^{n}(y_i - \bar{y})^2}} \quad (4)$$

The fits were made using the "polyfit" command in the Matlab software package (see Ref. [17] for details).

The $R^2$ value is a measure of how well the linear correlation fits the data. Nevertheless, from a practical point of view it is interesting to know how accurate these expressions are when they are used to predict surface tension values. In general, there is no direct relationship between the value of $R^2$ and the accuracy of the calculated surface tension. To calculate the AAD we first calculate the percentage deviation (PD) between the values obtained from the correlation by introducing the viscosity as input, $\sigma(\eta_i)$, and the data offered by NIST, $\sigma_i$, as follows:

$$\mathrm{PD}_i = 100\left(\sigma(T_i) - \sigma_i\right)/\sigma_i, \quad i = 1,2,...,N \tag{5}$$

where $N$ is the number of data considered for each fluid. We note that a positive $\mathrm{PD}_i$ value means that the model overestimates the accepted datum, whereas a negative $\mathrm{PD}_i$ value means that the model underestimates it. Then we calculated the average absolute percentage deviation for every fluid:

$$\mathrm{AAD} = \frac{\sum_{i=1}^{N}|\mathrm{PD}_i|}{N} \quad (\%) \tag{6}$$

It has to be borne in mind that, since AAD is a percentage, it is influenced by the higher individual PD values that are usually found at high temperatures, where the surface tension goes to zero and hence the relative deviations tend to increase considerably. In principle, a low AAD value does not mean that the PD values are low over the entire temperature range considered, so that high PDs may be found at particular temperatures, usually near the critical point.

In the following subsections we shall analyse the results considering both the $R^2$ and the AAD values, presenting several figures by way of illustrative examples. In those figures, not all the $N$ data used will be plotted for the sake of clarity in showing the behaviour of the correlations.

*a) Results for a wide temperature range*

The $R^2$ and AAD values for the 56 fluids, obtained using both the P and the MP correlations, Eqs. (1) and (3), are listed in Table 1. As can be seen, the $R^2$ values for the P correlation are greater than 0.998 for only 22 of the 56 fluids. They are below 0.98 for only seven fluids: R143a, carbon monoxide, deuterium oxide, hydrogen, oxygen, parahydrogen, and water. In the case of the MP correlation, the $R^2$ values are greater than 0.998 for 27 fluids, and no values are less than 0.98.

The AAD values for the P correlation are less than 2% only for R11, R113, R115, R245ca, and nonane, so that the accuracy of the use of the P expression in reproducing surface tension values is very limited. In particular, AADs greater than 20% are found for water, oxygen, and deuterium oxide, for which compounds the P model is therefore clearly inadequate in the wide temperature range considered.

In the following, we shall examine some examples of the different behaviours. Thus, an example of good performance and accuracy of the P expression is found for **R245ca**. Figure 1a shows that the P expression can adequately reproduce the data for this fluid with low absolute deviations, with AAD = 0.74%. In any case, as noted above, one needs to bear in mind that near the critical point (*i.e.*, for temperatures near $T_f$ = 443.86 K) the PD values increase significantly to even greater than 20%. The PD values are shown in Fig. 1b in which the values greater than 6% are not displayed.

Although for R245ca the MP expression is in principle not needed, the results are slightly improved by using an exponent value of $\phi$ = 1.0090 (see Table 1). The improvement at intermediate and low temperatures can be seen on the right-hand part of Fig. 1b.

When the P correlation is used for low n-alkanes, as was done by Queimada *et al.* [12], the results are inadequate, with the only exception being **nonane**. As can be seen in Fig. 2, for this fluid the linearity of Ln $\sigma$ versus the fluidity, $1/\eta$, is maintained over the whole temperature range. There is only a slight difference between the correlation proposed by Queimada *et al.*, which does not agree with the NIST at intermediate temperatures but does so at high temperatures, and that proposed here, which agrees with the NIST data in the high temperature range but deviates near the critical point. The difference can be appreciated in Fig. 3a, where only the values at high temperatures are shown. One can also see that the P and MP expression agree very well in this temperature range, so that the modification is not needed. As shown in Fig. 3b, the improvement of the MP expression with respect to the two P correlations is significant only at low temperatures, *i.e.*, near the triple point. The AAD value is only slightly reduced by using the MP expression instead of P. In general, the MP expression reduces the AAD values for all the n-alkanes, with the values obtained ranging from 1.7% for nonane to 4.30% for butane. If more accurate results are needed, the temperature range of validity has to be reduced. We shall consider this in the next subsection.

For some fluids, obtaining good linearity of Ln$\sigma$ versus $1/\eta$, *i.e.*, obtaining a high $R^2$ value, does not necessarily mean that the AAD value was low. A clear example is **R152a**. As can be observed from Table 1, the P expression led to a higher $R^2$ value than the MP one. When the surface tension is calculated from the viscosity values, the higher absolute deviations are appreciated only at low temperatures, as can be seen on the right-hand side of Fig. 4 (high viscosity values). Indeed, the P expression works better than the MP at these temperatures. Nevertheless, one observes in Table 1 that the MP correlation gives a lower AAD value. This is because AAD is a relative deviation and the PDs are higher at higher temperatures, *i.e.*, for the lower surface tension values (see the previous example in Fig. 1). In order to obtain better results for this and other fluids, it is necessary to reduce the temperature range of application of both the P and the MP correlations.

When the MP correlation is used, the surface tension data are reproduced with AADs below 2%

for 13 of the 56 fluids (only for 5 fluids in the case of the P correlation). The highest AAD value was obtained for R236ea (7.28%) despite the corresponding $R^2$ value not being low (0.997). This also means an improvement with respect to the P correlation, which gave AADs > 10% for 19 fluids (Table 1). For 34 fluids, such as R143a, carbon monoxide, deuterium oxide, hydrogen, oxygen, water, etc., the improvement is very significant. There are 7 fluids, however, for which the MP correlation is not significantly more accurate than the P one: R113, R123, R124, R141b, R218, R245ca (see Fig. 1), RC318, and nonane. Unfortunately, and contrary to the case of the ionic liquids studied by Ghatee *et al.* [10], the exponent $\phi$ does not take a fixed value, but varies from 0.8354 (dodecane) to 2.3792 (water).

Let us consider now two examples of the performance and accuracy of the MP expression. First, for argon the AAD is clearly reduced and the $R^2$ value is clearly increased. For this fluid, the data do not behave as exactly linear, although large deviations are only found at high temperatures, *i.e.*, near the critical point. Figure 5 shows the improvement of the MP correlation in reproducing the surface tension data when compared with the P one, especially at low temperatures (high viscosity values). In any case, some high PD values are found at high temperatures, so that the AAD obtained using the MP correlation is still high. As for other fluids, the temperature range of applicability has to be reduced in order for the model to work more accurately.

A clearer example of the improvement reached with the MP correlation is shown in Fig. 6, which is the case of hydrogen. For this fluid, the AAD is reduced from 18.95% with the P model, which is clearly inadequate, to only 2.89% with the MP.

Despite the good performance and accuracy of the MP correlation for several fluids, it cannot be applied with high accuracy over the whole temperature range from $T_0$ to $T_f$. If AAD values below 1% are needed in this temperature range, then it is adequate only for R115, R245ca, and neon. For the other fluids, the temperature range of applicability has to be reduced. This is considered in the next subsection.

*b) Accurate results for a reduced temperature range*

As indicate above, we have determined the temperature range at which the obtained AAD is below 1% for the two correlations considered. This temperature range starts at the $T_0$ value given in Table 1 and finishes at the new temperature values $T_{1p}$ and $T_{1m}$, for the P and MP models, respectively. These temperature values, expressed in reduced units, *i.e.*, divided by the critical point temperature, are given in Table 2. The new coefficients to be used in Eqs. (1) and (3) are given in Tables 3 and 4.

In the particular case of R245ca, both correlation models give AAD < 1% over the whole temperature range considered in the preceding subsection, so that the $T_{1p}$, $T_{1m}$, and $T_f$ values are identical. The same is the case when the MP model is applied to R115 and neon. For R113, the $T_{lp}$ and $T_{lm}$ values are identical and very close to $T_f$, which was to be expected since the AADs over the whole temperature range were both around 1.30%. For R32, the $T_{lp}$ and $T_{lm}$ values are identical, but clearly lower than $T_f$. This is because, for the whole temperature range, the AADs are 10.49% and 5.23%, respectively. The same is the case with hexane. In all the other cases, the reduced temperature range for the MP correlation is wider than that corresponding to the P correlation. In some cases, such as R11, R125, hydrogen sulfide, and octane for instance, the difference between the $T_{lp}$ and $T_{lm}$ values is very small, so that both correlation models are accurate in this reduced temperature range.

For the rest of the fluids, the improvement of the MP correlation model with respect to the P one is clearer. Let us consider the case of R152a, for instance. The behaviour of the correlations for the whole temperature range is presented in the data of Table 1 and in Fig. 4. Figure 7 shows that the MP correlation is very accurate from the lowest temperature to the point marked with an asterisk, which corresponds to the value at $T_{lm} = 0.91\ T_c$ and which is a wider temperature range than that corresponding to the $T_{lp} = 0.727\ T_c$ value. As is shown in the figure, when the new coefficients are used, the correlation cannot be extrapolated beyond the indicated temperature range.

There are 13 substances for which the reduced temperature range in which the MP correlation is very accurate is significantly larger than the corresponding P temperature range: R13, R123, R124, R141b, R142b, R218, R227ea, heptane, isobutane, propane, carbon monoxide, hydrogen, and parahydrogen. In the case of hydrogen, the corresponding surface tension and viscosity values for $T_{lp}$ and $T_{lm}$ are those shown in Fig. 6. As can be seen, the range of values in which the MP model can be very accurately applied is very wide.

The greatest difference between the reduced temperature ranges are found for R124, isobutane, and propane. For R124 we found a clear improvement at low temperatures (high surface tension and viscosity values) when the correlation models are used in the reduced temperature range. The behaviour at high temperatures is shown in Fig. 8, where only the viscosity values below 1.89 mPa s (or cP) are shown. As can be seen, the P expression is inadequate for this temperature range, whereas the MP correlation with the new coefficients is inaccurate only for viscosity values below 0.2 mPa s.

We consider finally, the cases of isobutane, deuterium oxide, oxygen, and water, for which the lowest $T_{lm}$ values are obtained even though their AAD values for the MP model over the whole temperature range are not high. The reason for these low values is that for these fluids the dependence of $\text{Ln}\sigma$ on $(1/\eta)^\phi$ is not linear at low temperatures, *i.e.*, near the triple point, and in our previous calculations the temperature range was reduced only at high temperatures. The inadequacy of the MP model at low temperatures is shown in Fig. 9 for the case of water. For this fluid the PD values are high at both the lowest (high viscosity values) and the highest (low viscosity values) temperatures. This means therefore that, for these four fluids, the reduced temperature range has not to exclude the lowest temperatures as well if the AAD is to be below 1%.

For these four fluids therefore, we considered what would be the case if the temperature range were cropped at either or both of the two ends for the MP model. This improved the results except in the case of isobutane. For this compound in particular, the reduced range constrained by the criterion of AAD < 1% was from 199.01 K to 391.64 K (the span thus being 192.63 K), whereas the uncropped temperature range was from 113.73 K (the triple point temperature) to $T_{1m}/T_c = 0.8017$ (see Table 2) i.e. $T_{1m} = 326.94$ K (the span thus being 213.21 K).

For the other three fluids, the new, and wider, the new reduced temperature ranges and the appropriate coefficients for Eq. (3) are listed in Table 5. In particular, for deuterium oxide, the previous reduced temperature range was from 276.97 K to 515.47 K, whereas the new one was from 355.86 K to 623.71 K, meaning an increase in the span of 29.25 K. For oxygen, the new range increased in span by just 7.51 K, and for water by 41.13 K.

As an example, in the case of water, the highest PD value is now around -5%. Fig. 10 shows the behaviour and accuracy of the proposed correlation in the reduced temperature range (that of Table 5) when applied to the whole temperature range (that in Table 1). As can be seen, the extrapolations lead to very inadequate results, so that the coefficients listed in Table 5 can be used only for the indicated temperature range.

## 4. Conclusions

We have studied the performance and accuracy of both the original Pelofsky relationship and the modified expression including an exponent over a wide range of temperatures for 56 non-ionic fluids of different kinds, including simple fluids (such as argon and carbon dioxide), simple alkanes, refrigerants, and some other substances such as water or ammonia. For each fluid, we checked the efficacy of the two correlations by calculating the linear coefficient $R^2$, and the accuracy of the correlations in reproducing the surface tension values from the viscosity data by calculating the PDs and AADs.

With the Pelofsky (P) correlation, the $R^2$ values were greater than 0.998 for only 22 of the 56 fluids, and were below 0.98 for seven fluids. The AAD values were less than 2% for only five fluids

(four refrigerants plus nonane), and greater than 20% for water, oxygen, and deuterium oxide. It can therefore be concluded that the performance and the accuracy of the P expression is very limited for the selected fluids and temperature ranges. This conclusion runs contrary to that of Pelofsky [13], the reason being that we have here considered different substances and temperature ranges.

With the use of the MP correlation, the surface tension data were reproduced with AADs below 2% for 13 of the 56 fluids considered, the poorest value being 7.3%. The improvement with the MP correlation was very significant for 34 fluids, but not significant for 7 fluids. Unfortunately, unlike the case of some ionic liquids, the exponent $\phi$ in the MP correlation did not take a fixed value, but varied from 0.8354 to 2.3792.

Since in both cases the AAD values were greater than 1% for most of the fluids, we delimited the temperature ranges to AAD < 1% for each correlation. The required coefficients were given for both the entire and the reduced temperature ranges for each expression. In all cases, the temperature range was reduced by excluding from consideration some high temperatures since at these temperatures the surface tension tends to zero so that the deviations increase significantly in percentage terms.

This reduction of the temperature range was not needed in the case of R245ca, for which both correlations were very accurate over the entire range. In the case of R115 and neon, this was the case only for the MP correlation, and for R113, R32, and hexane, there was no difference between using the P or MP correlation in the reduced temperature range. Thus, for those five fluids the use of the simpler P correlation has to be recommended. In all the other cases, the reduced temperature range for the MP correlation was wider than that for the P correlation, with the difference being particularly notable for 13 substances.

In the cases of water, oxygen, isobutane, and deuterium oxide, the dependence of Ln$\sigma$ on $(1/\eta)^{\phi}$ was not linear at low temperatures (near their triple points) when the MP correlation was used. We therefore delimited new reduced temperature ranges (for which the AADs values are below 1%) by excluding some high and low temperature values. With the exception of isobutane, this led to an increase in the span of the new ranges in which the MP correlation can be very accurately used.

Although it would be desirable, we have not been able to find any significant relationship between the performance and accuracy of the studied correlations and the molecular structure or properties of the different kinds of fluids. The main difficulty to achieve this is that we have considered just empirical correlations and we have studied different kinds of fluids. An extensive theoretical and computer simulation study on a particular kind of fluids would be needed in order to can establish a sound relationship between the behaviour of the viscosity and the surface tension.

**Acknowledgements**

The National Natural Science Foundation of China under Grants 10804061 and 11274200, the Natural Science Foundation of Shandong Province under Grant ZR2011AM017, and the foundations of QFNU and DUT have partially supported this work (X.L. and J.T.). It was also partially supported by Project FIS2010-16664 of the Ministerio de Economía y Competitividad of the Spanish Government, and the Gobierno de Extremadura and the European Union (FEDER) through project GR10045.

**FIGURE CAPTIONS**

Figure 1. Results for R245ca in the temperature range from $T_0$ (high values) to $T_f$ (near-to-zero values). a) Surface tension vs viscosity. Circles: selected NIST data; dashed line: P correlation; solid line: MP correlation. b) Percentage deviations (%). Dashed line: P correlation; solid line: MP correlation. Values near $T_f$ (viscosity near zero) are not shown, but they are around 22% for both correlations.

Figure 2. Values of Ln $\sigma$ versus $1/\eta$ (fluidity) for nonane in the temperature range from $T_0$ (low fluidity) to $T_f$ (high fluidity). Circles: selected NIST data; solid line: P correlation proposed here; Crosses: P correlation proposed by Queimada et al. [12].

Figure 3. Surface tension vs viscosity for nonane at the temperature range near the critical point (a), and near the triple point (b), respectively. Circles: selected NIST data; dashed line: P correlation; dash-dotted line: MP correlation. Crosses: Queimada et al. version of P correlation.

Figure 4. Surface tension vs viscosity for R152a in the temperature range from $T_0$ (high values) to $T_f$ (near-to-zero values). Circles: selected NIST data; dashed line: P correlation; dash-dotted line: MP correlation.

Figure 5. Surface tension values vs viscosity for argon in the $T_0$ (highest values) to $T_f$ (lowest values) range. Solid line: MP, Eq. (3); dashed line: P correlation, Eq. (1); circles: selected NIST data.

Figure 6. Surface tension values vs viscosity for hydrogen in the $T_0$ (highest values) to $T_f$ (lowest values) range. Solid line: MP, Eq. (3); dashed line: P correlation, Eq. (1); circles: selected NIST data. The asterisk, *, indicates that the MP correlation can be applied with AAD < 1% at higher surface tension values. The square, □, indicates the same but for the P correlation.

Figure 7. Values of Ln $\sigma$ vs $(1/\eta)^\phi$ for R152a. Solid line: MP correlation in the temperature range from $T_0 = 154.56$ K to $T_{1m} = 351.63$ K (marked with *); circles: selected NIST data. Here $\phi = 0.9498$ in Eq. (3).

Figure 8. Surface tension values vs viscosity for R124 at temperatures near the critical point. Circles: selected NIST data; dash-dotted line: MP correlation for the whole temperature range; solid line: MP correlation for the $T_0$-$T_{1m}$ range; dashed line: P correlation for the $T_0$ to $T_{1p}$ range; asterisk: data at $T_{1m}$; square point: data at $T_{1p}$.

Figure 9. Values of Ln $\sigma$ vs $(1/\eta)^\phi$ for water near the triple temperature. Solid line: MP correlation with the coefficients determined in the temperature range from $T_0$ (highest values) to $T_f$ (lowest values); circles: selected NIST data.

Figure 10. Surface tension values vs viscosity for water over the whole temperature range. Solid line: MP correlation with the coefficients determined in the reduced temperature range listed in Table 5; circles: selected NIST data.

**Table 1.** *The results of the P and MP correlations, Eqs. (1) and (3) respectively, over the whole temperature range from $T_0$ to $T_f$. The initial temperature $T_0$ is just the triple point temperature except for the substances marked with an asterisk, *, for which no data are available in NIST at low temperatures. The final temperature $T_f$ is the temperature nearest to the critical point given by the NIST Web Book. N denotes the number of data points, and $\phi$ is the value of the exponent in the MP expression. In the cases of ammonia and neon, the data presently available in NIST for the surface tension are not adequate, so those proposed by Mulero et al. [16] are used. Names of refrigerants are given below (+).*

| Substances | $T_o$ (K) | $T_c$ (K) | $T_f$ (K) | N | PELOFSKY | | Modified Pelofsky | | |
|---|---|---|---|---|---|---|---|---|---|
| | | | | | AAD | $R^2$ | $\phi$ | AAD | $R^2$ |
| REFRIGERANTS | | | | | | | | | |
| R11* | 198.15 | 471.11 | 466.48 | 175 | 1.97 | 0.9999 | 1.0366 | 1.42 | 0.9997 |
| R12* | 161.83 | 385.12 | 381.08 | 164 | 2.79 | 0.9997 | 1.0657 | 1.48 | 0.9997 |
| R13 | 92.00 | 302.00 | 300.95 | 200 | 5.82 | 0.9987 | 1.0798 | 4.75 | 0.9952 |
| R14 | 98.94 | 227.51 | 226.87 | 200 | 13.3 | 0.9873 | 1.4117 | 2.80 | 0.9973 |
| R22* | 158.84 | 369.30 | 366.76 | 165 | 4.45 | 0.9990 | 1.1140 | 1.98 | 0.9994 |
| R23 | 118.02 | 299.29 | 298.39 | 200 | 7.87 | 0.9978 | 1.1654 | 3.57 | 0.9972 |
| R32 | 136.34 | 351.25 | 350.18 | 200 | 10.5 | 0.9972 | 1.2467 | 5.23 | 0.9949 |
| R41 | 175.00 | 317.28 | 316.57 | 200 | 8.85 | 0.9961 | 1.2588 | 3.09 | 0.9988 |
| R113 | 236.93 | 487.21 | 482.2 | 197 | 1.30 | 0.9999 | 1.0043 | 1.29 | 0.9998 |
| R114* | 273.15 | 418.83 | 415.19 | 196 | 3.08 | 0.9995 | 1.1180 | 1.47 | 0.9998 |
| R115 | 173.76 | 353.10 | 348.62 | 196 | 1.12 | 0.9996 | 0.9595 | 0.37 | 1.0000 |
| R116 | 173.10 | 293.03 | 290.03 | 196 | 2.13 | 0.9999 | 1.0533 | 1.46 | 0.9998 |
| R123 | 166.00 | 456.83 | 455.38 | 200 | 7.93 | 0.9974 | 1.1231 | 6.65 | 0.9901 |
| R124* | 120.00 | 395.42 | 394.05 | 200 | 5.30 | 0.9931 | 1.0148 | 5.27 | 0.9919 |
| R125 | 172.52 | 339.17 | 338.34 | 200 | 5.54 | 0.9996 | 1.0981 | 4.01 | 0.9979 |
| R134a | 169.85 | 374.21 | 373.19 | 200 | 5.72 | 0.9995 | 1.1014 | 3.66 | 0.9983 |
| R141b | 169.68 | 477.50 | 475.96 | 200 | 5.88 | 0.9978 | 1.0735 | 5.19 | 0.9939 |
| R142b | 142.72 | 410.26 | 408.92 | 200 | 6.55 | 0.9987 | 1.1088 | 5.03 | 0.9945 |
| R143a | 161.34 | 345.86 | 344.93 | 200 | 19.1 | 0.9270 | 1.5960 | 1.80 | 0.9976 |
| R152a | 154.56 | 386.41 | 385.25 | 200 | 7.72 | 0.9988 | 1.1516 | 5.10 | 0.9946 |
| R218 | 125.45 | 345.02 | 343.92 | 200 | 5.19 | 0.9891 | 0.9578 | 4.95 | 0.9932 |
| R227ea | 146.35 | 375.95 | 374.80 | 200 | 4.21 | 0.9935 | 0.9225 | 3.17 | 0.9984 |
| R236ea* | 242.00 | 412.44 | 411.59 | 200 | 14.0 | 0.9949 | 1.3085 | 7.28 | 0.9970 |
| R236fa* | 186.08 | 398.07 | 396.98 | 194 | 6.83 | 0.9951 | 1.1601 | 1.42 | 0.9998 |
| R245ca* | 206.19 | 447.57 | 443.86 | 193 | 0.74 | 0.9999 | 1.0090 | 0.58 | 1.0000 |
| R245fa* | 200.00 | 427.20 | 426.06 | 200 | 7.17 | 0.9977 | 1.1399 | 3.32 | 0.9991 |
| RC318 | 233.35 | 388.38 | 387.60 | 200 | 3.47 | 0.9993 | 1.0478 | 3.11 | 0.9982 |
| HYDROCARBONS | | | | | | | | | |
| Butane | 134.90 | 425.12 | 423.67 | 200 | 6.32 | 0.9991 | 1.1098 | 4.30 | 0.9962 |
| Decane | 243.50 | 617.70 | 615.83 | 200 | 5.56 | 0.9991 | 1.1085 | 3.16 | 0.9977 |
| Dodecane | 263.60 | 658.10 | 656.13 | 200 | 8.70 | 0.9899 | 0.8354 | 2.66 | 0.9997 |
| Ethane | 90.352 | 305.33 | 304.26 | 200 | 10.9 | 0.9934 | 1.2597 | 4.09 | 0.9974 |
| Ethene | 103.99 | 282.35 | 281.46 | 200 | 10.2 | 0.9905 | 1.2638 | 2.60 | 0.9995 |
| Heptane | 182.55 | 540.13 | 538.34 | 200 | 5.75 | 0.9984 | 1.1242 | 3.11 | 0.9950 |
| Hexane | 177.83 | 507.82 | 506.17 | 200 | 8.25 | 0.9970 | 1.2141 | 2.92 | 0.9938 |
| Isobutane | 113.73 | 407.81 | 406.34 | 200 | 7.43 | 0.9969 | 1.1306 | 5.48 | 0.9893 |
| Methane | 90.694 | 190.56 | 190.06 | 200 | 12.6 | 0.9918 | 1.3690 | 3.96 | 0.9982 |
| Nonane | 219.70 | 594.55 | 592.68 | 200 | 1.94 | 0.9995 | 0.9806 | 1.70 | 0.9999 |
| Octane | 216.37 | 569.32 | 567.56 | 200 | 5.27 | 0.9994 | 1.0906 | 3.47 | 0.9983 |
| Pentane | 143.47 | 469.70 | 468.07 | 200 | 5.56 | 0.9979 | 1.1072 | 2.34 | 0.9995 |
| Propane | 85.48 | 369.82 | 368.40 | 200 | 6.96 | 0.9984 | 1.1213 | 4.79 | 0.9936 |
| Propene* | 100.00 | 365.57 | 364.24 | 200 | 11.2 | 0.9946 | 1.2660 | 4.30 | 0.9927 |
| 2-Methyl-Pentane* | 153.63 | 497.70 | 495.81 | 182 | 6.58 | 0.9989 | 1.1186 | 4.54 | 0.9948 |
| OTHERS | | | | | | | | | |
| Ammonia* | 198.64 | 405.40 | 402.25 | 195 | 12.5 | 0.9823 | 1.4684 | 2.78 | 0.9988 |
| Argon | 83.806 | 150.69 | 150.35 | 200 | 12.3 | 0.9935 | 1.3739 | 4.63 | 0.9980 |
| Carbon dioxide | 216.59 | 304.13 | 303.69 | 200 | 14.1 | 0.9912 | 1.5394 | 4.28 | 0.9971 |
| Carbon monoxide | 68.16 | 132.86 | 132.54 | 200 | 18.3 | 0.9495 | 1.7729 | 1.84 | 0.9998 |
| Deuterium oxide | 276.97 | 643.89 | 642.06 | 200 | 31.7 | 0.9011 | 2.1153 | 2.62 | 0.9926 |
| Hydrogen | 13.957 | 33.145 | 33.049 | 200 | 19.0 | 0.9590 | 1.9711 | 2.89 | 0.9982 |
| Hydrogen sulfide | 187.70 | 373.10 | 372.17 | 200 | 8.78 | 0.9987 | 1.2341 | 5.01 | 0.9958 |
| Krypton | 115.77 | 209.48 | 209.01 | 200 | 12.3 | 0.9874 | 1.4523 | 2.56 | 0.9989 |
| Neon* | 24.861 | 44.492 | 41.801 | 171 | 4.15 | 0.9943 | 1.4016 | 0.30 | 0.9999 |
| Nitrogen | 63.151 | 126.19 | 125.88 | 200 | 13.6 | 0.9887 | 1.4147 | 3.71 | 0.9975 |
| Oxygen | 54.361 | 154.58 | 154.08 | 200 | 22.4 | 0.9655 | 1.7456 | 4.00 | 0.9889 |
| Parahy- | 13.80 | 32.938 | 32.842 | 200 | 18.4 | 0.9611 | 1.9222 | 3.01 | 0.9984 |

| | | | | | | | | |
|---|---|---|---|---|---|---|---|---|
| drogen | | | | | | | | |
| Water | 273.16 | 647.10 | 645.23 | 200 | 36.9 | 0.8864 | 2.3792 | 3.45 | 0.9876 |
| Xenon | 161.40 | 289.73 | 289.09 | 200 | 9.21 | 0.9973 | 1.2901 | 3.95 | 0.9981 |

(+) Refrigerants: R11: Trichlorofluoromethane; R12: Dichlorodifluoromethane; R13: Chlorotrifluoromethane; R14: Tetrafluoromethane; R22: Chlorodifluoromethane; R23: Trifluoromethane (Fluoroform); R32: Difluoromethane; R41: Fluoromethane; R113: 1,1,2-Trichlorotrifluoroethane; R114: 1,2-Dichlorotetrafluoroethane; R115: Chloropentafluoroethane; R116: Hexafluoroethane; R123: 2,2-Dichloro-1,1,1-trifluoroethane; R124: 2-Chloro-1,1,1,2-tetrafluoroethane; R125: Pentafluoroethane; R134a: 1,1,1,2-Tetrafluoroethane; R141b: 1,1-Dichloro-1-fluoroethane; R142b: 1-Chloro-1,1-difluoroethane; R143a: 1,1,1-Trifluoroethane; R152a: 1,1-Difluoroethane; R218: Octafluoropropane; R227ea: 1,1,1,2,3,3,3-Heptafluoropropane; R236ea: 1,1,1,2,3,3-Hexafluoropropane; R236fa: 1,1,1,3,3,3-Hexafluoropropane; R245ca: 1,1,2,2,3-Pentafluoropropane; R245fa: 1,1,1,3,3-Pentafluoropropane; RC318: Octafluorocyclobutane (Perfluorocyclobutane).

**Table 2.** Reduced values for the temperatures $T_{1p}$ and $T_{1m}$ which satisfy the requirement of the AAD of the P or the MP expressions, respectively, being less than 1%.

| Substances | $T_{1p}/T_c$ | $T_{1m}/T_c$ | Substances | $T_{1p}/T_c$ | $T_{1m}/T_c$ |
|---|---|---|---|---|---|
| REFRIGERANTS | | | Decane | 0.9364 | 0.9520 |
| R11* | 0.9804 | 0.9840 | Dodecane | 0.7872 | 0.9101 |
| R12* | 0.9651 | 0.9825 | Ethane | 0.8240 | 0.9402 |
| R13 | 0.5967 | 0.8992 | Ethene | 0.7347 | 0.9842 |
| R14 | 0.8050 | 0.9294 | Heptane | 0.5664 | 0.8378 |
| R22* | 0.9416 | 0.9760 | Hexane | 0.8473 | 0.8473 |
| R23 | 0.8880 | 0.9395 | Isobutane | 0.4628 | 0.8017 |
| R32 | 0.9021 | 0.9021 | Methane | 0.8349 | 0.9607 |
| R41 | 0.9193 | 0.9731 | Nonane | 0.8928 | 0.9905 |
| R113 | 0.9846 | 0.9846 | Octane | 0.9628 | 0.9690 |
| R114 | 0.9704 | 0.9861 | Pentane | 0.8437 | 0.9861 |
| R115 | 0.9568 | 0.9873 | Propane | 0.4657 | 0.8616 |
| R116 | 0.9795 | 0.9836 | Propene | 0.8293 | 0.8402 |
| R123 | 0.6180 | 0.8822 | 2-Methyl-Pentane* | 0.6581 | 0.8940 |
| R124 | 0.4532 | 0.8538 | OTHERS | | |
| R125 | 0.9558 | 0.9631 | Ammonia | 0.7618 | 0.9612 |
| R134a | 0.9645 | 0.9672 | Argon | 0.9046 | 0.9667 |
| R141b | 0.5906 | 0.8970 | Carbon dioxide | 0.9280 | 0.9669 |
| R142b | 0.6479 | 0.9055 | Carbon monoxide | 0.6810 | 0.9781 |
| R143a | 0.6319 | 0.8719 | Deuterium oxide | 0.7065 | 0.8006 |
| R152a | 0.7270 | 0.9100 | Hydrogen | 0.6990 | 0.9595 |
| R218 | 0.5195 | 0.8820 | Hydrogen sulfide | 0.9280 | 0.9329 |
| R227ea | 0.6733 | 0.9603 | Krypton | 0.8367 | 0.9776 |
| R236ea | 0.9380 | 0.9711 | Neon | 0.8119 | 0.9395 |
| R236fa* | 0.8435 | 0.9945 | Nitrogen | 0.8376 | 0.9551 |
| R245ca* | 0.9917 | 0.9917 | Oxygen | 0.7374 | 0.7764 |
| R245fa | 0.9335 | 0.9787 | Parahydrogen | 0.6950 | 0.9652 |
| RC318 | 0.8423 | 0.9660 | Water | 0.6995 | 0.7775 |
| HYDROCARBONS | | | Xenon | 0.9424 | 0.9646 |
| Butane | 0.8874 | 0.9110 | | | |

**Table 3.** Proposed coefficients for the Pelofsky (P) expression in the two temperature ranges.

| Substances | $T_0 \sim T_f$ | | $T_0 \sim T_{1p}$ | | Substances | $T_0 \sim T_f$ | | $T_0 \sim T_{1p}$ | |
|---|---|---|---|---|---|---|---|---|---|
| | $B$ | $\ln A$ | $B$ | $\ln A$ | | $B$ | $\ln A$ | $B$ | $\ln A$ |
| REFRIGERANTS | | | | | Decane | -0.2298 | -3.4336 | -0.2156 | -3.4926 |
| R11* | -0.2895 | -3.2983 | -0.2859 | -3.3106 | Dodecane | -0.1915 | -3.6688 | -0.2384 | -3.5248 |
| R12* | -0.2717 | -3.3423 | -0.2650 | -3.3680 | Ethane | -0.1590 | -3.1939 | -0.1343 | -3.3395 |
| R13 | -0.2600 | -3.4151 | -0.2736 | -3.4322 | Ethene | -0.1765 | -3.1474 | -0.1425 | -3.3472 |
| R14 | -0.2594 | -3.2856 | -0.2027 | -3.5484 | Heptane | -0.2038 | -3.3442 | -0.2090 | -3.3846 |
| R22* | -0.2703 | -3.2199 | -0.2562 | -3.2737 | Hexane | -0.2075 | -3.2914 | -0.1816 | -3.4003 |
| R23 | -0.2593 | -3.1792 | -0.2338 | -3.2753 | Isobutane | -0.1876 | -3.4040 | -0.2464 | -3.4038 |
| R32 | -0.2493 | -2.9476 | -0.2184 | -3.0785 | Methane | -0.1190 | -3.2191 | -0.0988 | -3.4761 |
| R41 | -0.2562 | -3.1396 | -0.2268 | -3.2988 | Nonane | -0.2070 | -3.4821 | -0.2090 | -3.4791 |
| R113 | -0.3069 | -3.5844 | -0.3050 | -3.5903 | Octane | -0.2046 | -3.4143 | -0.1938 | -3.4649 |
| R114 | -0.2630 | -3.5365 | -0.2546 | -3.5845 | Pentane | -0.1945 | -3.2939 | -0.1793 | -3.3603 |
| R115 | -0.2777 | -3.7096 | -0.2793 | -3.7044 | Propane | -0.1751 | -3.2599 | -0.1890 | -3.2937 |
| R116 | -0.2828 | -3.8170 | -0.2783 | -3.8363 | Propene | -0.1923 | -3.1327 | -0.1622 | -3.2672 |
| R123 | -0.3164 | -3.3865 | -0.3541 | -3.3921 | 2-Methyl-Pentane* | -0.1981 | -3.3341 | -0.1932 | -3.3825 |
| R124 | -0.2999 | -3.4705 | -0.5145 | -3.3905 | OTHERS | | | | |
| R125 | -0.2944 | -3.5481 | -0.2763 | -3.6159 | Ammonia | -0.1815 | -2.5874 | -0.1335 | -2.8638 |
| R134a | -0.2821 | -3.4035 | -0.2660 | -3.4616 | Argon | -0.2045 | -3.4878 | -0.1738 | -3.7225 |
| R141b | -0.2637 | -3.3221 | -0.2997 | -3.3159 | Carbon dioxide | -0.2662 | -2.8563 | -0.2178 | -3.2049 |
| R142b | -0.2618 | -3.3513 | -0.2658 | -3.3839 | Carbon monoxide | -0.2488 | -3.1097 | -0.1305 | -3.9015 |
| R143a | -0.2181 | -3.6892 | -0.0530 | -4.2079 | Deuterium oxide | -0.2322 | -2.0373 | -0.1039 | -2.5283 |
| R152a | -0.2232 | -3.3109 | -0.2111 | -3.3783 | Hydrogen | -0.0210 | -4.6779 | -0.0128 | -5.2869 |
| R218 | -0.3011 | -3.8048 | -0.5264 | -3.6809 | Hydrogen sulfide | -0.2467 | -2.7668 | -0.2219 | -2.9081 |
| R227ea | -0.2915 | -3.6854 | -0.3652 | -3.5876 | Krypton | -0.2970 | -3.2642 | -0.2332 | -3.5761 |
| R236ea | -0.3456 | -3.4878 | -0.2978 | -3.6557 | Neon | -0.0996 | -4.4386 | -0.0874 | -4.5829 |
| R236fa* | -0.2953 | -3.6046 | -0.2565 | -3.7090 | Nitrogen | -0.1396 | -3.7692 | -0.1110 | -4.0340 |
| R245ca* | -0.3038 | -3.5391 | -0.3038 | -3.5391 | Oxygen | -0.2120 | -3.2627 | -0.1383 | -3.6375 |
| R245fa | -0.3242 | -3.4639 | -0.2967 | -3.5419 | Parahydrogen | -0.0207 | -4.7077 | -0.0128 | -5.2973 |
| RC318 | -0.3551 | -3.8008 | -0.3578 | -3.8113 | Water | -0.2167 | -1.9875 | -0.0917 | -2.5197 |
| HYDROCARBONS | | | | | Xenon | -0.3992 | -3.0815 | -0.3579 | -3.2543 |
| Butane | -0.1838 | -3.2955 | -0.1702 | -3.3631 | | | | | |

**Table 4.** Proposed coefficients for the modified Pelofsky (MP) expression in the two temperature ranges.

| Substances | $T_0 \sim T_f$ | | | $T_0 \sim T_{1m}$ | | |
|---|---|---|---|---|---|---|
| | $D$ | $\ln C$ | $\phi$ | $D$ | $\ln C$ | $\phi$ |
| REFRIGERANTS | | | | | | |
| R11* | -0.2617 | -3.3425 | 1.0366 | -0.2691 | -3.3350 | 1.0235 |
| R12* | -0.2254 | -3.4243 | 1.0657 | -0.2330 | -3.4154 | 1.0509 |
| R13 | -0.2069 | -3.4979 | 1.0798 | -0.2866 | -3.4221 | 0.9304 |
| R14 | -0.0753 | -3.7868 | 1.4117 | -0.1036 | -3.7195 | 1.2853 |
| R22* | -0.1950 | -3.3601 | 1.1140 | -0.2132 | -3.3369 | 1.0753 |
| R23 | -0.1598 | -3.3675 | 1.1654 | -0.1951 | -3.3257 | 1.0783 |
| R32 | -0.1203 | -3.2243 | 1.2467 | -0.2116 | -3.0888 | 1.0123 |
| R41 | -0.1143 | -3.5288 | 1.2588 | -0.1521 | -3.4455 | 1.1509 |
| R113 | -0.3034 | -3.5895 | 1.0043 | -0.3102 | -3.5830 | 0.9936 |
| R114* | -0.1826 | -3.7415 | 1.1180 | -0.1959 | -3.7133 | 1.0919 |
| R115 | -0.3112 | -3.6552 | 0.9595 | -0.3112 | -3.6552 | 0.9595 |
| R116 | -0.2423 | -3.8975 | 1.0533 | -0.2556 | -3.8767 | 1.0315 |
| R123 | -0.2294 | -3.5050 | 1.1231 | -0.4092 | -3.3409 | 0.8338 |
| R124* | -0.2885 | -3.4837 | 1.0148 | -0.4100 | -3.3959 | 0.8235 |
| R125 | -0.2223 | -3.6795 | 1.0981 | -0.2957 | -3.5853 | 0.9750 |
| R134a | -0.2114 | -3.5245 | 1.1014 | -0.2561 | -3.4753 | 1.0152 |
| R141b | -0.2146 | -3.3950 | 1.0735 | -0.3149 | -3.3020 | 0.8937 |
| R142b | -0.1930 | -3.4589 | 1.1088 | -0.2837 | -3.3689 | 0.9310 |
| R143a | -0.0371 | -4.1844 | 1.5960 | -0.0239 | -4.2248 | 1.7929 |
| R152a | -0.1417 | -3.4687 | 1.1516 | -0.2302 | -3.3575 | 0.9498 |
| R218 | -0.3367 | -3.7633 | 0.9578 | -0.4630 | -3.6685 | 0.7883 |
| R227ea | -0.3624 | -3.5898 | 0.9225 | -0.4227 | -3.5358 | 0.8496 |
| R236ea* | -0.1505 | -3.8642 | 1.3085 | -0.2347 | -3.7443 | 1.1056 |
| R236fa* | -0.1896 | -3.7775 | 1.1601 | -0.0189 | -4.2573 | 1.1490 |
| R245ca* | -0.2967 | -3.5483 | 1.0090 | -0.2967 | -3.5483 | 1.0090 |
| R245fa* | -0.2233 | -3.6151 | 1.1399 | -0.2535 | -3.5885 | 1.0761 |
| RC318 | -0.3112 | -3.8745 | 1.0478 | -0.3984 | -3.7652 | 0.9382 |
| HYDROCARBONS | | | | | | |
| Butane | -0.1291 | -3.4138 | 1.1098 | -0.1819 | -3.3443 | 0.9768 |
| Decane | -0.1656 | -3.5556 | 1.1085 | -0.1966 | -3.5193 | 1.0362 |
| Dodecane | -0.3320 | -3.4255 | 0.8354 | -0.3259 | -3.4367 | 0.8438 |
| Ethane | -0.0665 | -3.4607 | 1.2597 | -0.0927 | -3.4105 | 1.1400 |
| Ethene | -0.0725 | -3.4778 | 1.2638 | -0.0838 | -3.4522 | 1.2124 |
| Heptane | -0.1387 | -3.4715 | 1.1242 | -0.2103 | -3.3859 | 0.9487 |
| Hexane | -0.1068 | -3.5115 | 1.2141 | -0.1839 | -3.3973 | 0.9949 |
| Isobutane | -0.1250 | -3.5180 | 1.1306 | -0.2431 | -3.3984 | 0.8466 |
| Methane | -0.0280 | -3.7761 | 1.3690 | -0.0446 | -3.6814 | 1.2344 |
| Nonane | -0.2201 | -3.4584 | 0.9806 | -0.2293 | -3.4498 | 0.9635 |
| Octane | -0.1543 | -3.5152 | 1.0906 | -0.1837 | -3.4805 | 1.0196 |
| Pentane | -0.1388 | -3.4077 | 1.1072 | -0.1511 | -3.3940 | 1.0729 |
| Propane | -0.1190 | -3.3694 | 1.1213 | -0.1808 | -3.3024 | 0.9521 |
| Propene* | -0.0838 | -3.3772 | 1.2660 | -0.1481 | -3.2851 | 1.0377 |
| 2-Methyl-Pentane* | -0.1369 | -3.4542 | 1.1186 | -0.2081 | -3.3685 | 0.9493 |
| OTHERS | | | | | | |
| Ammonia* | -0.0405 | -3.0685 | 1.4684 | -0.0541 | -3.0250 | 1.3629 |
| Argon | -0.0578 | -4.0553 | 1.3739 | -0.1004 | -3.9118 | 1.1859 |
| Carbon dioxide | -0.0470 | -3.7395 | 1.5394 | -0.1004 | -3.5196 | 1.2708 |
| Carbon monoxide | -0.0209 | -4.1773 | 1.7729 | -0.0261 | -4.1407 | 1.6934 |
| Deuterium oxide | -0.0107 | -2.6894 | 2.1153 | -0.0403 | -2.6058 | 1.5040 |
| Hydrogen | $-1.0126 \cdot 10^{-4}$ | -5.7133 | 1.9711 | $-2.4422 \cdot 10^{-4}$ | -5.6547 | 1.7985 |
| Hydrogen sulfide | -0.1178 | -3.1241 | 1.2341 | -0.2374 | -2.8759 | 0.9771 |
| Krypton | -0.0781 | -3.8973 | 1.4523 | -0.0984 | -3.8419 | 1.3601 |
| Neon* | -0.0226 | -4.8773 | 1.4016 | -0.0226 | -4.8773 | 1.4016 |
| Nitrogen | -0.0307 | -4.2974 | 1.4147 | -0.0471 | -4.2247 | 1.2799 |
| Oxygen | -0.0220 | -3.8948 | 1.7456 | -0.0880 | -3.7170 | 1.1918 |
| Parahydrogen | $-1.2950 \cdot 10^{-4}$ | -5.7055 | 1.9222 | $-2.9430 \cdot 10^{-4}$ | -5.6502 | 1.7612 |
| Water | -0.0045 | -2.7079 | 2.3792 | -0.0346 | -2.5943 | 1.4904 |
| Xenon | -0.1786 | -3.5775 | 1.2901 | -0.2751 | -3.3967 | 1.1073 |

**Table 5.** New reduced temperature ranges and coefficients for the MP correlation, Eq. (3), for which AADs < 1% are found. $N$ is the new number of data considered in the selected temperature range.

| Substance | $T_{range}$ | D | Ln C | $\phi$ | N |
|---|---|---|---|---|---|
| Deuterium Oxide | 355.86K-623.71K | -7.0627 10$^{-3}$ | -2.7652 | 2.2644 | 147 |
| Oxygen | 75.908K-149.07K | -0.0180 | -3.9834 | 1.8015 | 147 |
| Water | 355.43K-626.53K | -3.4139 10$^{-3}$ | -2.7784 | 2.4709 | 146 |